\documentclass[11pt,a4paper]{emulateapj}
\usepackage{verbatim}
\usepackage{epsfig}
\usepackage{amsmath}
\usepackage{natbib}

\begin{document}

\title{Toward the Detection of Exoplanet Transits with Polarimetry}
\author{Sloane J. Wiktorowicz\altaffilmark{1} \& Gregory P. Laughlin}
\affil{Department of Astronomy and Astrophysics, University of California, Santa Cruz, CA 95064}
\email{sloanew@ucolick.org}

\altaffiltext{1}{NASA Sagan Fellow}

\date{\today}

\begin{abstract}

In contrast to photometric transits, whose peak signal occurs at mid-transit due to occultation of the brightest region of the disk, polarimetric transits provide a signal upon ingress and egress due to occultation of the polarized stellar limb. Limb polarization, the bright corollary to limb darkening, arises from the $90^\circ$ scattering angle and low optical depth experienced by photons at the limb. In addition to the ratio $R_{\rm p} / R_*$, the amplitude of a polarimetric transit is expected to be controlled by the strength and width of the stellar limb polarization profile, which depend on the scattering-to-total opacity ratio at the stellar limb. We present a short list of the systems providing the highest expected signal-to-noise ratio for detection of this effect, and we draw particular attention to HD 80606b. This planet is spin/orbit misaligned, has a three-hour ingress, and has a bright parent star, which make it an attractive target. We report on test observations of an HD 80606b ingress with the POLISH2 polarimeter at the Lick Observatory Shane 3-m telescope. We conclude that unmodeled telescope systematic effects prevented polarimetric detection of this event. We outline a roadmap for further refinements of exoplanet polarimetry, whose eventual success will require a further factor of ten reduction in systematic noise.

\end{abstract}

\section{Introduction}

Polarimetry provides a unique, but not yet mainstream, technique for directly extracting information from exoplanets. Observed from afar, starlight scattered by a planetary atmosphere or surface will be polarized and will modulate through the full range in orbital phase angles (e.g., \citealp{Seager2000, Stam2004}). Repeated observations of the full phase curve require short-period exoplanets, where direct light from the host star dominates. Since scattered light detection of these exoplanets requires instrumental accuracy of $\la 10^{-5}$, the science of exoplanetary polarimetry has not yet matured \citep{Wiktorowicz14}. Unlike transit photometry and transmission spectroscopy, polarimetry is not biased towards edge-on systems. However, transiting exoplanets allow the most rigorous demonstration of detection, because variability due to the instrument, telescope, sky, interstellar medium, and host star may be assessed during successive secondary eclipses. \\

An exoplanet also induces a polarimetric signature during transit. Observations of both the Sun {\citep{Faurobert2001, Faurobert2003} and Algol \citep{Kemp1983} have verified the prediction of stellar limb polarization \citep{Chandrasekhar1946a, Chandrasekhar1946b}. Limb darkening, due to low optical depth and to photon diffusion preferentially in the plane of the sky at the limb, ensures that most photons scattered at the limb do not reach the observer. However, these also ensure that the relatively few limb photons scattered toward the observer will be preferentially polarized tangent to the limb. \\

Thus, while the largest {\it photometric} transit signal occurs at mid-transit, the largest {\it polarimetric} signal occurs during ingress/egress as the exoplanet occults the stellar limb \citep{Carciofi2005, Kostogryz2011}. This breaks the symmetry of limb-integrated, tangential polarization and introduces net radial polarization typically at the part-per-million level. For spherical stars and exoplanets, the polarization position angle during transit is parallel to the line joining the star-planet centers. Indeed, this position angle rotates with a sense determined by the stellar hemisphere transited. Polarimetric transits and phase curves independently constrain the planet's longitude of the ascending node $\Omega$, which is not measurable from photometric or radial velocity observations. Therefore, mutually consistent constraints on $\Omega$ are valuable in validating the technique of exoplanet polarimetry. \\

Starspots may also introduce a polarimetric signal at the limb, but the projected area of the spot vanishes. Thus, while \citet{Berdy2011a} model that one large starspot ($1\%$ of the area of the stellar disk) provides a polarimetric amplitude of $3 \times 10^{-6}$, the expected amplitude is an order of magnitude lower due to projection effects. In addition, repeated transits sample different starspot conditions, which further mitigate their effect. \\

In \S \ref{model1}, we outline a model permitting evaluation of the expected in-transit stellar polarization, and we discuss its expected variation with respect to key parameters such as $R_{\rm p} / R_*$. We describe unique science afforded by transit polarimetry in \S \ref{section_apps}, and section \S \ref{section_results} summarizes our HD 80606b test observations. We provide prospects for improved data quality in \S \ref{section_future} and present concluding remarks in \S \ref{section_conclusion}. \\

\section{Polarimetric Transit Model}
\label{model1}
\subsection{Basic Equations}
\label{model}

By adopting a numerical approach similar to \citet[hereafter CM05]{Carciofi2005}, we model the expected polarization due to an exoplanet transiting a sunlike star as a function of $R_{\rm p} / R_*$. Since disk-integrated linear polarization from a spherical, featureless star is zero, and that polarization position angle is tangent to the limb, net polarization during transit is equal to the fractional circumference occulted by the exoplanet multiplied by the polarization and limb darkened stellar intensity at each radius step along the stellar disk. For exoplanets where $R_{\rm p} / R_* \ll 1$,

\begin{eqnarray}
P(t) & = & \int_{r_{\rm p} (t) - R_{\rm p} / R_*}^{r_{\rm p} (t) + R_{\rm p} / R_*} \! \frac{C(r,t)}{2 \pi r} I(r) P_*(r) \, {\rm d}r. \label{model}
\end{eqnarray}

\noindent Here, $P(t)$ is instantaneous transit polarization, $r_{\rm p} (t)$ is the instantaneous position of the exoplanet center, $C(r,t)$ is the instantaneous path length along the stellar circumference at $r$ that is occulted by the exoplanet, $I(r)$ is the limb-darkened stellar intensity, and $P_*(r)$ is the stellar polarization at radius $r$. For a spherical exoplanet,

\begin{eqnarray}
C(r,t) \approx 2 \sqrt{ (R_{\rm p} / R_*)^2 - [ r - r_{\rm p} (t) ]^2}.
\end{eqnarray}

\noindent Given the nearly constant apparent velocity of the exoplanet across the disk of the star even for eccentric exoplanets \cite[hereafter H10]{Hebrard2010},

\begin{eqnarray}
r_{\rm p}(t) & = & \sqrt{b^2 + 4 [ ( 1 + R_{\rm p} / R_*)^2 - b^2 ] (t / T_{14})^2}.
\label{rvst}
\end{eqnarray}

\noindent Here, $b$ is the transit impact parameter, $t$ is the time since mid-transit, and $T_{14}$ is the transit duration. \\

Limb-darkened stellar intensity is given by

\begin{eqnarray}
I(\mu) & = & 1 - c_1 (1 - \mu) (2 - \mu) / 2 + c_2 (1 - \mu) \mu / 2
\end{eqnarray}

\noindent \citep{Brown01, barnes03}, where \linebreak $\mu = \cos{\theta} = \cos({ \arcsin{r} })$ is the angle between the observer's line of sight and the normal to the stellar surface. Stellar polarization is modeled as

\begin{eqnarray}
P_*(\mu) & = & P_1 \left( \frac{1 - \mu^2}{1 + k \mu} \right)
\end{eqnarray}

\noindent \citep{Fluri1999}, where $P_1$ represents the degree of polarization at the limb and $k$ is the inverse of the profile width, which determines how rapidly polarization decreases from the limb to the stellar disk center. \\

In addition to modeling the degree of polarization $P(t)$ during a transit (Equation \ref{model}), we also model polarization position angle. Since stellar polarization is tangential to the limb, the polarization position angle during transit, $\Theta (t)$, is equal to the plane-of-sky position angle of the planetary center with respect to the stellar center (Equation \ref{rvst}):

\begin{eqnarray}
\Theta (t) & = & \pm \arctan \left[ \frac{2t}{T_{14}} \sqrt{ \left( \frac{(1 + R_{\rm p} / R_*}{b} \right)^2 - 1} \, \right] + C, ~
\end{eqnarray}

\noindent where $C$ is a constant. Thus, polarization position angle rotates monotonically with time during the transit, and the sense of rotation (clockwise or counterclockwise) depends on the stellar hemisphere that is transited. \\

\begin{figure}
\includegraphics[scale=0.59]{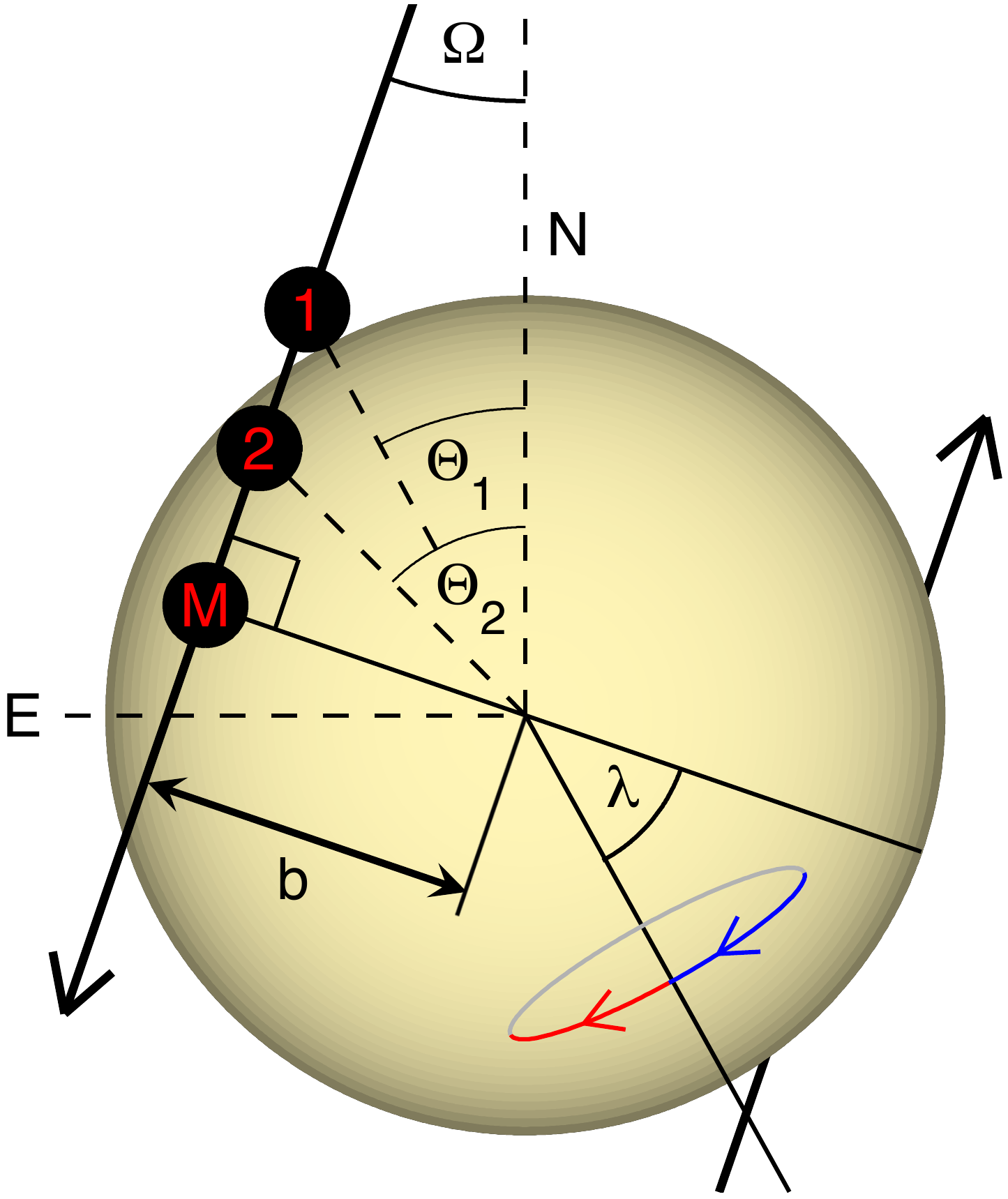}
\caption{Hypothetical, plane-of-sky transit geometry. Exoplanet locations are shown for first and second contacts and mid-transit. The spin axis of the star is shown, where $\lambda = 42^\circ \pm 8^\circ$ (H10), and the blue- and red-shifted stellar hemispheres are indicated. The entire system may be rotated by $180^\circ$ due to the inherent ambiguity in linear polarimetry.}
\label{pazoom_synth}
\end{figure}

Rotation of the observed polarization position angle between first and second contacts is given by the following:

\begin{eqnarray}
\Theta_2 - \Theta_1 & = & \arccos \left( \frac{b}{1 + R_{\rm p} / R_*} \right) \\
& - &  \arccos \left( \frac{b}{1 - R_{\rm p} / R_*} \right) \nonumber,
\end{eqnarray}

\noindent Therefore, HD 80606b is expected to cause polarization rotation by $| \Theta_2 - \Theta_1 | = 16\, \fdg 21 \pm 0\, \fdg 76$ between first and second contacts given $b = 0.808 \pm 0.007$ and $R_{\rm p} / R_* = 0.1001 \pm 0.0006$ (H10). While polarimetric measurement of $\Theta_1$ and $\Theta_2$ may in principle constrain $b$, it is unlikely that this will improve upon the accuracy of photometrically-derived values. \\

Uniquely, polarimetry provides an estimate of the planet's longitude of the ascending node $\Omega$ via

\begin{eqnarray}
\Omega & = & \Theta_1 + \arccos \left( \frac{b}{1 + R_{\rm p} / R_*} \right) - 90^\circ \\
 & = & \Theta_2 + \arccos \left( \frac{b}{1 - R_{\rm p} / R_*} \right) - 90^\circ.
\end{eqnarray}

\noindent Figure \ref{pazoom_synth} shows a hypothetical plane-of-sky geometry of the HD 80606 system. The planet's path is indicated by thick black lines, where transit and occultation lines may be reversed due to the inherent $\pm 180^\circ$ ambiguity in linear polarization\label{polamb}. The Rossiter-McLaughlin effect imparts a significant blue-shift of the system near third contact (H10), so the exoplanet must transit the red-shifted limb of the star during egress. The common proper motion companion HD 80607 \citep{Mugrauer2006}, which may be responsible for maintaining HD 80606b's high eccentricity via the Kozai mechanism \citep{Wu03}, lies nearly due East of the HD 80606 system. Estimation of $\Omega$ assumes a circular cross section for both star and exoplanet, and we briefly discuss the effect of stellar oblateness in \S \ref{section_obl}. \\

\subsection{Scaling with Planet/Star Radius Ratio}

\begin{figure}
\includegraphics[scale=0.49]{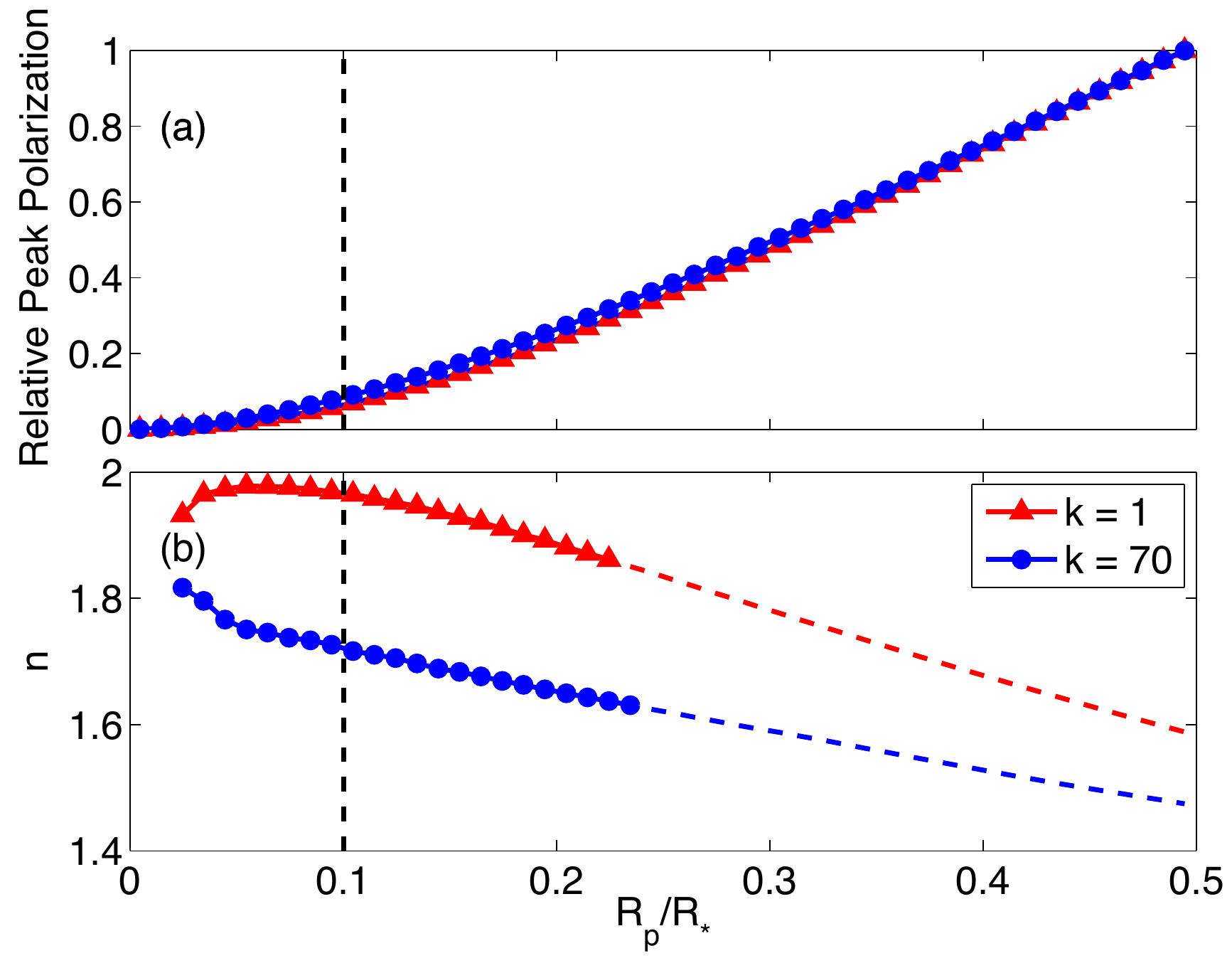}
\caption{(a). Relative polarimetric amplitude of a transit versus $R_{\rm p} / R_*$ for a narrow, sunlike limb polarization profile (blue circles) and for a broad, Mira-like profile (red triangles). The vertical, black dashed line represents $R_{\rm p} / R_*$ for HD 80606b. (b) The best-fit, power-law exponent for polarimetric amplitude bounded by $[0, R_{\rm p} / R_*]$. Blue and red dashed lines indicate $> 3 \sigma$ departure from a power-law behavior.}
\label{polamp}
\end{figure}

We reproduce the results of CM05 and confirm that polarimetric amplitude during transit is insensitive to impact parameter $b$ for non-grazing transits. Figure \ref{polamp}a shows the relative polarimetric amplitude of a transit versus $R_{\rm p} / R_*$ for $R_{\rm p} / R_* < 0.5$ and $b = 0$. We adopt HD 80606 limb darkening coefficients \citep[$c_1 = 0.742$ and $c_2 = 0.458$]{Roberts2013}, as well as 460 nm solar limb polarization profile coefficients \citep[$P_1 = 0.87 \%$ and $k = 70$]{Faurobert2001}, because limb polarization is difficult to measure for most stars. \\

Given the strong dependence of limb polarization profile and wavelength of observation on the polarimetric transit amplitude, we scale to peak polarization $\Delta P = \max [ P(t) ]$ during transit in Figure \ref{polamp}a. The power-law exponent $n$ in polarimetric amplitude, $\Delta P \propto (R_{\rm p} / R_*)^n$, is shown in Figure \ref{polamp}b for the best fit to numerical results in the interval $[0, R_{\rm p} / R_*]$. For hot Jupiters around sunlike stars, $R_{\rm p} / R_* \approx 0.1$ and polarimetric amplitude scales roughly as $\Delta P \propto (R_{\rm p} / R_*)^{1.75}$. This is only slightly shallower than in photometry, where $\Delta F \propto (R_{\rm p} / R_*)^2$. The polarimetric power-law exponent decreases with increasing planet/star radius ratio up to $R_{\rm p} / R_* \approx 0.25$, where a $\chi^2$ analysis rejects a power-law fit at the $3 \sigma$ level. For the sake of illustration, however, the dashed lines in Figure \ref{polamp}b show the best fit power-law exponent for $R_{\rm p} / R_* \ga 0.25$. \\

To explore the variation produced by markedly different stellar atmospheres, we analyze polarimetric amplitudes for an extremely broad, $k = 1$ limb polarization profile similar to the prediction for red giants such as Mira (\citealt{Harrington1969}, CM05). This has virtually no effect on the relative polarimetric amplitude versus $R_{\rm p} / R_*$ (Figure \ref{polamp}a), although it increases the power-law exponent nearly to $n = 2$, which is the value for a photometric transit (Figure \ref{polamp}b). Intuitively, in the limit of an infinitely sharp limb polarization profile, transit polarization scales with the stellar circumference ($n = 1$). In contrast, the polarimetric amplitude from an extremely broad profile scales with the ratio of areas ($n = 2$). Late-type stars are expected to harbor a stronger degree of limb polarization $P_1$ than that of sunlike stars, because molecular condensation in their cooler atmospheres increases the contribution of scattering opacity \citep{Harrington1969}. This suggests that super Earths orbiting late-type stars may present important targets for future polarimetric transit observations. \\

For observations where sensitivity scales with photon statistics, such as those in this study (\S \ref{section_results}), the SNR of transit polarization scales with the following product: 1) the square root of the total number of stellar photons detected during ingress/egress, and 2) polarization amplitude $\propto (R_{\rm p} / R_*)^n$. For a given system, the {\it photometric} transit duration scales with $R_*$, while the {\it polarimetric} duration scales with $R_{\rm p}$. Thus, for a typical hot Jupiter with a photometric transit duration of a few hours, the polarimetric transit signal will only last roughly one-half hour at each ingress/egress. Planetary ingress/egress duration (the time between first and second or third and fourth contacts) is not always published; however, it may be estimated from the following:

\begin{eqnarray}
T_{12} = T_{14} \left( \frac{1}{2} - \sqrt{\frac{1}{4} - \frac{R_{\rm p}/R_*}{(1 + R_{\rm p}/R_*)^2 - b^2}} \right)
\end{eqnarray}

\noindent From ingress/egress duration, $R_{\rm p} / R_*$, and apparent stellar magnitude $m$ in the wavelength band of observation, the relative SNR of transit polarization scales with the following:

\begin{eqnarray}
{\rm SNR} & \propto & 10^{-0.2m} (R_{\rm p}/R_*)^n \sqrt{T_{12} }.
\end{eqnarray}

\noindent For $k = 70$, Figure \ref{polamp}b suggests $n = -0.77 (R_{\rm p}/R_*) + 1.80$. \\

Table \ref{targets} lists the ten most favorable systems for polarimetric transit detection. Here, SNR is calculated for a single transit and does not account for the number of transits observable over a given time period, which modifies SNR by a factor of $T_{\rm orb}^{-1/2}$. Additional modifications due to stellar spectral type are expected (Kostogryz \& Berdyugina, in preparation). The hot Jupiters HD 189733b and HD 209458b are highly favorable systems for polarimetric transit detection due to their bright host stars. The highly eccentric HD 80606b has significant radial velocity at transit, which occurs nearly six days after periastron. This lengthens its photometric transit duration to $\sim 12$ hours and ingress/egress to $\sim 3$ hours (H10). Therefore, the second brightest, spin/orbit misaligned system known ($\lambda = 42^\circ \pm 8^\circ$, H10), HD 80606b is also one of the best candidates for observation of a polarimetric transit. The next most favorable target is WASP-33b, which is also the brightest misaligned system currently known ($\lambda = -108\, \fdg02 \pm 0\, \fdg52$, \citealp{Collier2010}). The significant oblateness of this host star may be detectable with transit polarimetry. \\

\begin{deluxetable}{l c c c c c}
\tabletypesize{\normalsize}
\tablecaption{Most Favorable Polarimetric Transits}
\tablewidth{0pt}
\tablehead{
\colhead{Exoplanet} & \colhead{SNR} & \colhead{$R_{\rm p}/R_*$} & \colhead{$m_V$} & \colhead{$T_{12}$ (min)} & \colhead{$T_{\rm orb}$ (day)}}
\startdata
HD 189733b	& 1		& 0.155 & 7.7 & 24 & 2.2 \\
HD 209458b	& 0.65	& 0.121 & 7.7 & 26 & 3.5 \\
HD 80606b	& 0.62	& 0.106 & 9.1 & 145 & 111.4 \\
WASP-33b	& 0.29	& 0.107 & 8.3 & 16 & 1.2 \\
HD 17156b	& 0.28	& 0.073 & 8.2 & 53 & 21.2 \\
WASP-34b	& 0.26	& 0.112 & 10.3 & 63 & 4.3 \\
HAT-P-30b	& 0.20	& 0.113 & 10.4 & 43 & 2.8 \\
HAT-P-17b	& 0.19	& 0.124 & 10.5 & 29 & 10.3 \\
WASP-79b	& 0.19	& 0.107 & 10.0 & 31 & 3.7 \\
HAT-P-1b		& 0.18	& 0.112 & 10.3 & 32 & 4.5 \\
\enddata
\label{targets}
\end{deluxetable}

\section{Unique Applications}
\label{section_apps}

\subsection{Stellar Oblateness}
\label{section_obl}
Since the rotation of polarization position angle during transit is determined by the morphology of the stellar limb, stellar oblateness may be directly detectable. \citet{Bailey2010} demonstrate that the large degree of polarization of Regulus, with respect to stars with similar parallax, is consistent with its oblate nature as determined by interferometry \citep{McAlister2005}. However, intrinsic stellar polarization is contaminated by interstellar polarization, which is dependent on heliocentric distance, galactic longitude, and mean ISM grain size along the line of sight. Therefore, stellar polarization is likely only a qualitative estimator of stellar oblateness. We detail a purely geometric estimator that is independent of interstellar polarization. \\

For spin/orbit alignment ($\lambda = 0^\circ$), it is trivial to show that the position angle of the stellar limb occulted by the planet upon ingress/egress is given by

\begin{eqnarray}
\Theta_* = \pm \arctan \left( \frac{(1-f) \sqrt{1-b^2}}{b} \right) + C.
\end{eqnarray}

\noindent Here, stellar oblateness $f = 1 - R_{\rm pole} / R_{\rm equator}$, where $R_{\rm pole}$ and $R_{\rm equator}$ indicate the stellar polar and equatorial radii, respectively. Occultation of polarization tangent to the limb introduces polarization perpendicular to the limb (i.e., perpendicular to $\Theta_*$). Regardless of oblateness, the rotation of polarization position angle between ingress and egress peaks for $b = 1 / \sqrt{2}$, and this rotation is $90^\circ$ for a spherical star (Figure \ref{oblate}a). The difference in maximum rotation between an oblate and spherical star is nearly linear (Figure \ref{oblate}b) and is given by

\begin{eqnarray}
\Delta \Theta_* = 90^\circ - 2 \arctan{(1-f)} \approx 58^\circ f.
\end{eqnarray}

\noindent The maximum rotation difference due to stellar oblateness reaches $\Delta \Theta_* \approx 1^\circ$ for $f = 0.02$, which corresponds to the rapidly rotating WASP-33 ($v \sin I_* = 86.29 \pm 0.31$ km/s: \citealp{Collier2010}). Since this is also a high SNR system (Table \ref{targets}), we suggest that it be observed with high priority. In contrast, the low stellar $v \sin I_* = 1.7 \pm 0.3$ km/s for HD 80606 (H10) implies a stellar oblateness of only $f = 8 \times 10^{-6}$, which is undetectable with transit polarimetry. \\

\begin{figure}
\includegraphics[scale=0.69]{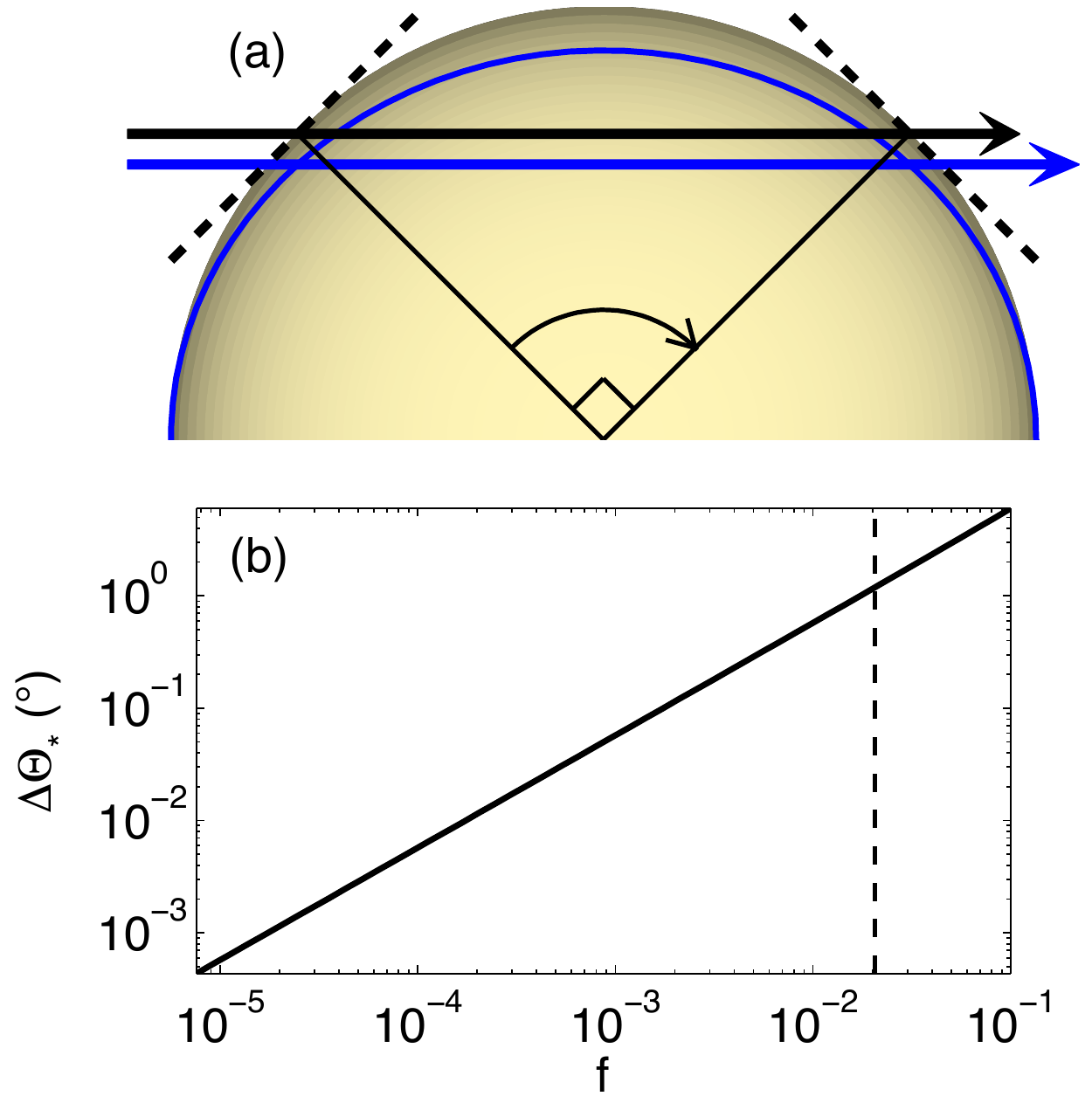}
\caption{(a) Transits of spherical and oblate stars. For a spherical star with $b = 1 / \sqrt 2$ and $\lambda = 0^\circ$, polarization rotates $90^\circ$ between ingress and egress (dashed lines indicate tangential limb polarization and solid lines indicate radial transit polarization). Even for the same $b$, the magnitude of rotation decreases for an oblate star. For example, the strongly oblate limb in blue, with $f = 0.1$, only provides $84^\circ$ of polarization rotation. (b) The difference in ingress/egress rotation between spherical and oblate stars versus $f$. Expected HD 80606 and WASP-33 stellar oblatenesses are $f = 8 \times 10^{-6}$ (left plot bound) and $f = 0.02$ (dashed line), respectively. \\}
\label{oblate}
\end{figure}

\subsection{Planetary Dynamics}
\label{section_coplan}

For highly eccentric exoplanets orbiting one component of a binary system, constraint of $\Omega$ determines the orientation of the planet's orbit in the plane of the sky. The addition of astrometric and radial velocity observations of the binary's orbit provides the full space motion of the exoplanet with respect to the binary companion. The efficacy of the Kozai mechanism in perturbation of the planet's eccentricity may then be assessed. As another example, measurement of the interior structure of HAT-P-13b may be obtained if the mutual inclination of the planetary pair can be determined \citep{Batygin2009, Mardling2010}. The combination of radial velocity, transit, polarimetric, and astrometric data may generate a fully constrained solution. \\

\section{Observations}
\label{section_results}

The POLISH2 polarimeter at the Lick Observatory Shane 3-m telescope uses two photoelastic modulators to simultaneously measure Stokes $I$, $Q$, $U$, and $V$ (see \citealt{Wiktorowicz2008, Wiktorowicz2009} for the prototype POLISH). Using this system, we obtain $B$ band observations during a transit ingress of HD 80606b (first contact 06:17 UT, February 1, 2013: H10; \citealt{Shporer2010, Roberts2013}). We also obtain 4.9 hours of control data during the previous night to complement the 7.1 hour observation during transit. Each minute of on-star observations is followed by 30 seconds on a sky field for sky subtraction. Every six minutes, we transition between HD 80606 and HD 80607. HD 80607 is a common proper motion companion that is likely bound to HD 80606, $\sim 20$ arcsec away \citep{Mugrauer2006}. Both stars are of similar spectral type and apparent magnitude, but HD 80607 is not known to harbor an exoplanet. Sky-subtracted, Stokes $Q$ and $U$ measurement uncertainties in six-minute bins demonstrate that sensitivity on each star scales as $\sigma_{Q,U} = 1.87 / \sqrt{N}$ for $N$ detected photons (Figure \ref{photscaling}). Measurement sensitivity therefore scales with photon statistics but lies $87\%$ from the photon limit. \\

\begin{figure}
\includegraphics[scale=0.49]{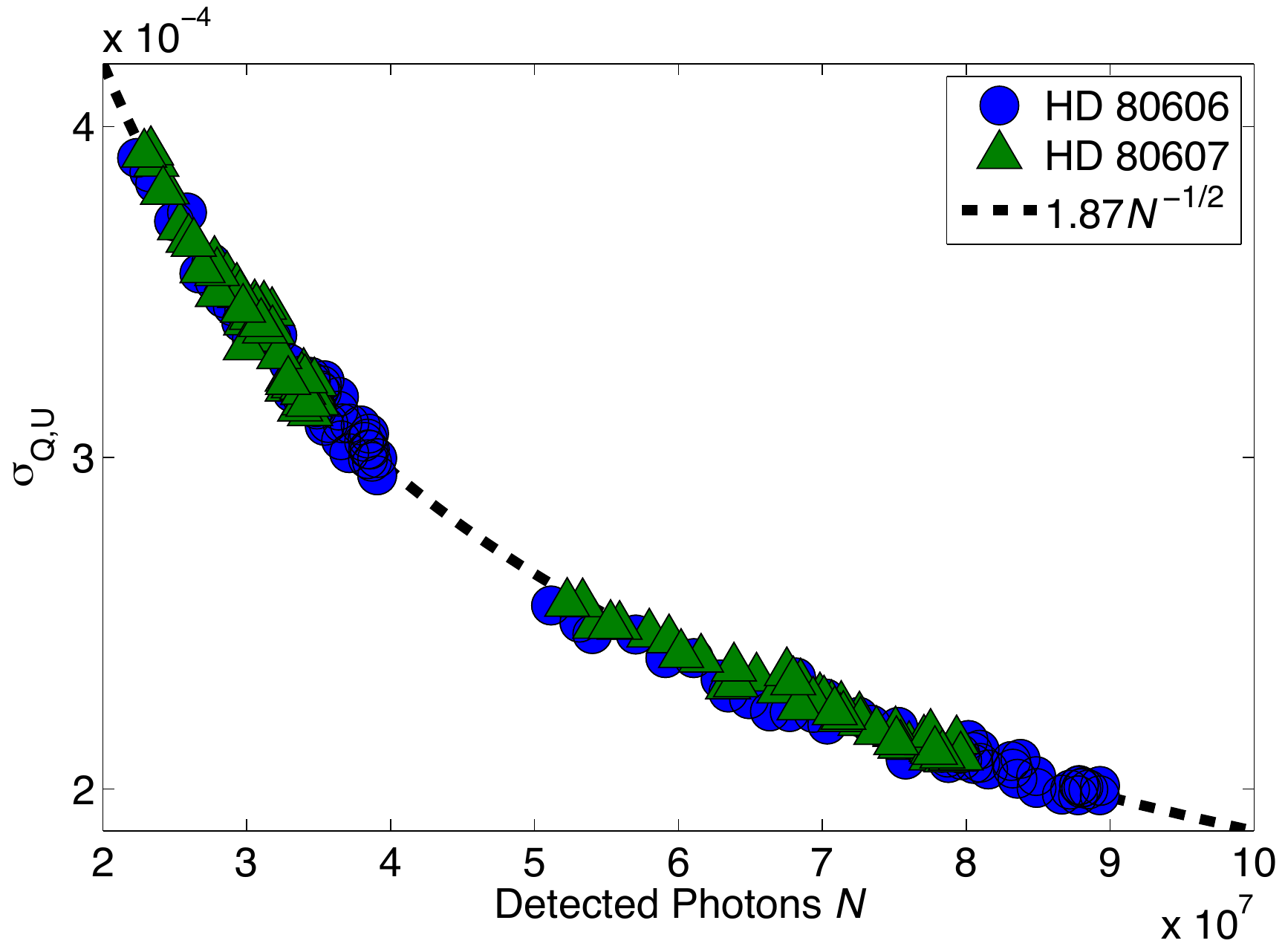}
\caption{The standard error of the mean within all Stokes $Q$, $U$ bins comprising HD 80606 and HD 80607 observations. Known instrumental gain factors are applied to Stokes $I$ measurements to determine the number of detected photons $N$.}
\label{photscaling}
\end{figure}

Figure \ref{transzoom} shows the measured difference in Stokes $Q$, $U$, $P = \sqrt{Q^2 + U^2}$ (degree of linear polarization), and $\Theta = 1/2 \arctan(U / Q)$ (position angle of linear polarization) between HD 80606 and HD 80607. Out-of-transit, control data are shown at left, and transit data are at right. Three red, vertical lines during transit indicate first and second contacts and mid-transit. Current data quality does not support conclusive detection, as a constant fit to observations cannot be rejected ($\chi^2_\nu = 0.53$, $\nu = 48$). A model with transit-derived properties is shown in blue (Figure \ref{transzoom}); however, an amplitude match requires an unphysically large limb polarization ($P_1 > 50\%$) that even a pure scattering atmosphere cannot provide, as $\max(P_1) \sim 12\%$ \citep{Chandrasekhar1946a, Fluri1999}. \\

\begin{figure}
\includegraphics[scale=0.66]{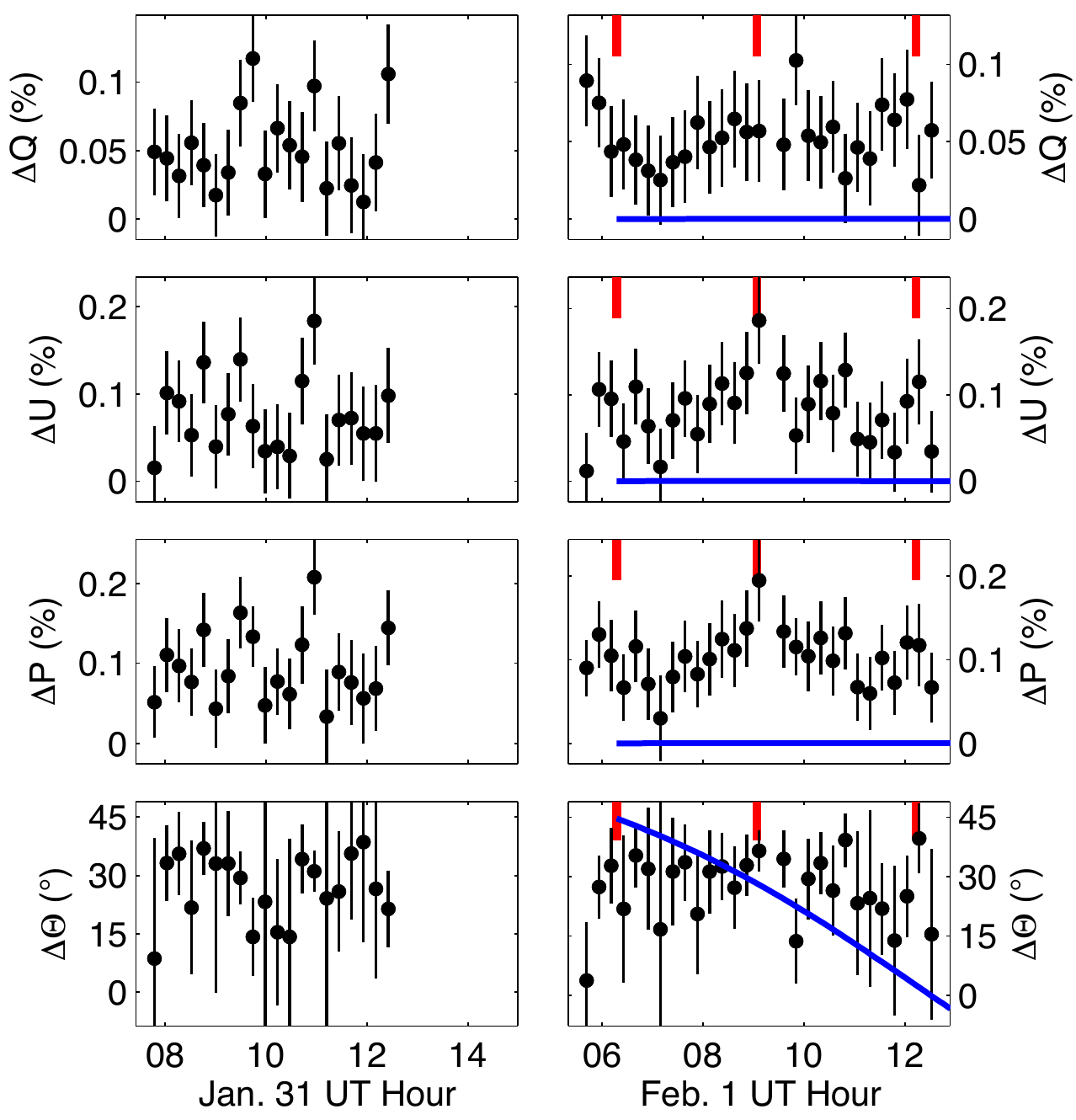}
\caption{Polarimetric difference between HD 80606 and HD 80607 for each night individually. The left column represents the control night, and the right represents the transit night. The time duration in each plot is identical. Times of first and second contacts and mid-transit are shown by red lines. Blue curves indicate a model with transit-derived properties; however, this model requires an unphysically large limb polarization.}
\label{transzoom}
\end{figure}

\section{Ongoing Control of Systematic Effects}
\label{section_future}

Accuracy in nearly all cutting-edge exoplanetary detection and characterization is dominated by non-Gaussian systematic effects, not photon noise. For example, systematic effects of the Earth-trailing {\it Spitzer Space Telescope} require reduction by two orders of magnitude to uncover published results (e.g., \citealt{Knutson2012, Demory2013}). Design of POLISH2 for minimal systematic effects requires their suppression by only a factor of a few to achieve accuracy necessary for exoplanet science, even at an historic telescope overlooking a major metropolitan area. Therefore, an advanced polarimeter at a larger, modern telescope (e.g., Keck Observatory) is likely to deliver conclusive detection of a polarimetric transit. \\

We foresee promising avenues for improvement of POLISH2 measurement accuracy:

-- Unpublished observations of polarized stars suggest that telescope flexure during long observations causes a measurable change in polarization induced by the telescope. The amplitude of this effect appears correlated with stellar polarization. While run-averaged polarizations of HD 80606 and HD 806067 are consistent with zero ($(Q, U)_{\rm HD80606} = (-1.1 \pm 3.0, +3.8 \pm 3.3) \times 10^{-5}$ and $(Q, U)_{\rm HD80607} = (+2.9 \pm 3.2, +2.9 \pm 3.5) \times 10^{-5}$), further modeling of calibrator star data is likely to improve measurement accuracy.

-- Since POLISH2's photoelastic modulators are resonant devices, thermal and humidity variations affect instantaneous retardance. This requires a time-variable scaling factor to be applied to polarization measurements. Since we continuously measure the resonant frequency of each PEM, we expect that a physically-motivated model relating retardance drift to resonant frequency will improve measurement accuracy.

\section{Conclusion}
\label{section_conclusion}

We revisit the modeling of polarimetric transits and demonstrate that rotation of observed polarization position angle allows constraint of the planetary longitude of the ascending node $\Omega$. We determine the ten most favorable transiting systems for detection of this effect. The highest priority targets are two well-studied hot Jupiters around bright stars (HD 189733b and HD 209458b), the highly eccentric, misaligned HD 80606b, and the retrograde hot Jupiter orbiting the oblate star WASP-33. The departure of measured polarization position angle rotation with respect to a spherical star may allow a direct measurement of WASP-33 stellar oblateness. Constraint of $\Omega$ for HD 80606b may assess the feasibility of the Kozai mechanism in maintaining the planet's extreme eccentricity. For the eccentric HAT-P-13 system, polarimetric transit observations of planet b may be combined with astrometry of the non-transiting brown dwarf c to assess coplanarity. This result will enable constraint of $k_2$ for the transiting planet, which would constitute a direct physical probe of an exoplanetary interior. \\

We attempt to observe a single polarimetric ingress of HD 80606b using the POLISH2 polarimeter at the Lick Observatory Shane 3-m telescope, but unmodeled systematic effects currently preclude detection. However, ongoing suppression of systematic effects at the $10^{-5}$ level, due to polarimetric drift of both the telescope and instrument, are expected to refine high sensitivity polarimetry into a reliably powerful technique in exoplanetary science on 8-10m class, ground-based telescopes. \\

\acknowledgments
This work was performed (in part) under contract with the California Institute of Technology (Caltech) funded by NASA through the Sagan Fellowship Program executed by the NASA Exoplanet Science Institute. SJW would like to acknowledge support from the NASA Origins of Solar Systems program through grant NNX13AF63G. \\
 
 {\it Facilities:} \facility{Shane (POLISH2)}.


\begin{thebibliography}{}
\expandafter\ifx\csname natexlab\endcsname\relax\def\natexlab#1{#1}\fi

\bibitem[{{Bailey} {et~al.}(2010){Bailey}, {Lucas}, \& {Hough}}]{Bailey2010}
{Bailey}, J., {Lucas}, P.~W., \& {Hough}, J.~H. 2010, \mnras, 405, 2570

\bibitem[{{Barnes} \& {Fortney}(2003)}]{barnes03}
{Barnes}, J.~W., \& {Fortney}, J.~J. 2003, \apj, 588, 545

\bibitem[{{Batygin} {et~al.}(2009){Batygin}, {Bodenheimer}, \&
  {Laughlin}}]{Batygin2009}
{Batygin}, K., {Bodenheimer}, P., \& {Laughlin}, G. 2009, \apjl, 704, L49

\bibitem[{{Berdyugina} {et~al.}(2011){Berdyugina}, {Berdyugin}, {Fluri}, \&
  {Piirola}}]{Berdy2011a}
{Berdyugina}, S.~V., {Berdyugin}, A.~V., {Fluri}, D.~M., \& {Piirola}, V. 2011,
  \apjl, 728, L6

\bibitem[{{Brown} {et~al.}(2001){Brown}, {Charbonneau}, {Gilliland}, {Noyes},
  \& {Burrows}}]{Brown01}
{Brown}, T.~M., {Charbonneau}, D., {Gilliland}, R.~L., {Noyes}, R.~W., \&
  {Burrows}, A. 2001, \apj, 552, 699

\bibitem[{{Carciofi} \& {Magalh{\~a}es}(2005)}]{Carciofi2005}
{Carciofi}, A.~C., \& {Magalh{\~a}es}, A.~M. 2005, \apj, 635, 570

\bibitem[{{Chandrasekhar}(1946{\natexlab{a}})}]{Chandrasekhar1946a}
{Chandrasekhar}, S. 1946{\natexlab{a}}, \apj, 103, 351

\bibitem[{{Chandrasekhar}(1946{\natexlab{b}})}]{Chandrasekhar1946b}
---. 1946{\natexlab{b}}, \apj, 104, 110

\bibitem[{{Collier Cameron} {et~al.}(2010){Collier Cameron}, {Guenther},
  {Smalley}, {McDonald}, {Hebb}, {Andersen}, {Augusteijn}, {Barros}, {Brown},
  {Cochran}, {Endl}, {Fossey}, {Hartmann}, {Maxted}, {Pollacco}, {Skillen},
  {Telting}, {Waldmann}, \& {West}}]{Collier2010}
{Collier Cameron}, A., {Guenther}, E., {Smalley}, B., {et~al.} 2010, \mnras,
  407, 507

\bibitem[{{Demory} {et~al.}(2013){Demory}, {de Wit}, {Lewis}, {Fortney},
  {Zsom}, {Seager}, {Knutson}, {Heng}, {Madhusudhan}, {Gillon}, {Barclay},
  {Desert}, {Parmentier}, \& {Cowan}}]{Demory2013}
{Demory}, B.-O., {de Wit}, J., {Lewis}, N., {et~al.} 2013, \apjl, 776, L25

\bibitem[{{Faurobert} \& {Arnaud}(2003)}]{Faurobert2003}
{Faurobert}, M., \& {Arnaud}, J. 2003, \aap, 412, 555

\bibitem[{{Faurobert} {et~al.}(2001){Faurobert}, {Arnaud}, {Vigneau}, \&
  {Frisch}}]{Faurobert2001}
{Faurobert}, M., {Arnaud}, J., {Vigneau}, J., \& {Frisch}, H. 2001, \aap, 378,
  627

\bibitem[{{Fluri} \& {Stenflo}(1999)}]{Fluri1999}
{Fluri}, D.~M., \& {Stenflo}, J.~O. 1999, \aap, 341, 902

\bibitem[{{Harrington}(1969)}]{Harrington1969}
{Harrington}, J.~P. 1969, \apjl, 3, 165

\bibitem[{{H{\'e}brard} {et~al.}(2010){H{\'e}brard}, {D{\'e}sert},
  {D{\'{\i}}az}, {Boisse}, {Bouchy}, {Lecavelier Des Etangs}, {Moutou},
  {Ehrenreich}, {Arnold}, {Bonfils}, {Delfosse}, {Desort}, {Eggenberger},
  {Forveille}, {Gregorio}, {Lagrange}, {Lovis}, {Pepe}, {Perrier}, {Pont},
  {Queloz}, {Santerne}, {Santos}, {S{\'e}gransan}, {Sing}, {Udry}, \&
  {Vidal-Madjar}}]{Hebrard2010}
{H{\'e}brard}, G., {D{\'e}sert}, J.-M., {D{\'{\i}}az}, R.~F., {et~al.} 2010,
  \aap, 516, A95

\bibitem[{{Kemp} {et~al.}(1983){Kemp}, {Henson}, {Barbour}, {Kraus}, \&
  {Collins}}]{Kemp1983}
{Kemp}, J.~C., {Henson}, G.~D., {Barbour}, M.~S., {Kraus}, D.~J., \& {Collins},
  II, G.~W. 1983, \apjl, 273, L85

\bibitem[{{Knutson} {et~al.}(2012){Knutson}, {Lewis}, {Fortney}, {Burrows},
  {Showman}, {Cowan}, {Agol}, {Aigrain}, {Charbonneau}, {Deming}, {D{\'e}sert},
  {Henry}, {Langton}, \& {Laughlin}}]{Knutson2012}
{Knutson}, H.~A., {Lewis}, N., {Fortney}, J.~J., {et~al.} 2012, \apj, 754, 22

\bibitem[{{Kostogryz} {et~al.}(2011){Kostogryz}, {Yakobchuk}, {Morozhenko}, \&
  {Vid'Machenko}}]{Kostogryz2011}
{Kostogryz}, N.~M., {Yakobchuk}, T.~M., {Morozhenko}, O.~V., \& {Vid'Machenko},
  A.~P. 2011, \mnras, 415, 695

\bibitem[{{Mardling}(2010)}]{Mardling2010}
{Mardling}, R.~A. 2010, \mnras, 407, 1048

\bibitem[{{McAlister} {et~al.}(2005){McAlister}, {ten Brummelaar}, {Gies},
  {Huang}, {Bagnuolo}, {Shure}, {Sturmann}, {Sturmann}, {Turner}, {Taylor},
  {Berger}, {Baines}, {Grundstrom}, {Ogden}, {Ridgway}, \& {van
  Belle}}]{McAlister2005}
{McAlister}, H.~A., {ten Brummelaar}, T.~A., {Gies}, D.~R., {et~al.} 2005,
  \apj, 628, 439

\bibitem[{{Mugrauer} {et~al.}(2006){Mugrauer}, {Neuh{\"a}user}, {Mazeh},
  {Guenther}, {Fern{\'a}ndez}, \& {Broeg}}]{Mugrauer2006}
{Mugrauer}, M., {Neuh{\"a}user}, R., {Mazeh}, T., {et~al.} 2006, Astronomische
  Nachrichten, 327, 321

\bibitem[{{Roberts} {et~al.}(2013){Roberts}, {Barnes}, {Rowe}, \&
  {Fortney}}]{Roberts2013}
{Roberts}, J.~E., {Barnes}, J.~W., {Rowe}, J.~F., \& {Fortney}, J.~J. 2013,
  \apj, 762, 55

\bibitem[{{Seager} {et~al.}(2000){Seager}, {Whitney}, \&
  {Sasselov}}]{Seager2000}
{Seager}, S., {Whitney}, B.~A., \& {Sasselov}, D.~D. 2000, \apj, 540, 504

\bibitem[{{Shporer} {et~al.}(2010){Shporer}, {Winn}, {Dreizler}, {Col{\'o}n},
  {Wood-Vasey}, {Choi}, {Morley}, {Moutou}, {Welsh}, {Pollaco}, {Starkey},
  {Adams}, {Barros}, {Bouchy}, {Cabrera-Lavers}, {Cerutti}, {Coban},
  {Costello}, {Deeg}, {D{\'{\i}}az}, {Esquerdo}, {Fernandez}, {Fleming},
  {Ford}, {Fulton}, {Good}, {H{\'e}brard}, {Holman}, {Hunt}, {Kadakia},
  {Lander}, {Lockhart}, {Mazeh}, {Morehead}, {Nelson}, {Nortmann}, {Reyes},
  {Roebuck}, {Rudy}, {Ruth}, {Simpson}, {Vincent}, {Weaver}, \&
  {Xie}}]{Shporer2010}
{Shporer}, A., {Winn}, J.~N., {Dreizler}, S., {et~al.} 2010, \apj, 722, 880

\bibitem[{{Stam} {et~al.}(2004){Stam}, {Hovenier}, \& {Waters}}]{Stam2004}
{Stam}, D.~M., {Hovenier}, J.~W., \& {Waters}, L.~B.~F.~M. 2004, \aap, 428, 663

\bibitem[{{Wiktorowicz}(2009)}]{Wiktorowicz2009}
{Wiktorowicz}, S.~J. 2009, \apj, 696, 1116

\bibitem[{{Wiktorowicz} \& {Matthews}(2008)}]{Wiktorowicz2008}
{Wiktorowicz}, S.~J., \& {Matthews}, K. 2008, \pasp, 120, 1282

\bibitem[{{Wiktorowicz} \& {Stam}(2015)}]{Wiktorowicz14}
{Wiktorowicz}, S.~J., \& {Stam}, D.~M. 2015, in Polarization of Stars and
  Planetary Systems, ed. L.~{Kolokolova}, J.~{Hough}, \& A.-C.
  {Levasseur-Regourd} (Cambridge University Press), in press

\bibitem[{{Wu} \& {Murray}(2003)}]{Wu03}
{Wu}, Y., \& {Murray}, N. 2003, \apj, 589, 605

\end{thebibliography}

 \end{document}